\documentclass[useAMS,usenatbib]{mn2e}
\usepackage{graphicx,amsmath,multirow,amssymb}
\usepackage{subfig}
\usepackage{natbib}
\newcommand{\comment}[1]{}

\def\simgt{\lower.5ex\hbox{$\; \buildrel > \over \sim \;$}}
\def\simlt{\lower.5ex\hbox{$\; \buildrel < \over \sim \;$}}

\title[O enrichment in C-rich PNe]{Probing O-enrichment in C-rich dust planetary nebulae}
\author[Garc\'{\i}a--Hern\'andez et al.]{D. A.~Garc\'{\i}a-Hern\'andez$^{1,2}$,
P. Ventura$^3$, G. Delgado-Inglada$^4$ F. Dell'Agli$^{3,5}$, 
\newauthor
M. Di Criscienzo$^1$, A. Yag\"ue$^{3,1,2}$\\
$^{1}$Instituto de Astrof\'{\i}sica de Canarias, E-38205 La Laguna, Tenerife, Spain \\
$^{2}$Departamento de Astrof\'{\i}sica, Universidad de La Laguna (ULL), E-38206 La Laguna, Tenerife, Spain\\
$^3$INAF -- Osservatorio Astronomico di Roma, Via Frascati 33, 00040, Monte Porzio Catone (RM), Italy \\
$^4$Instituto de Astronom\'{\i}a, Universidad Nacional Aut\'onoma de M\'exico, Apdo. Postal 70264,04510, M\'exico D. F., M\'exico\\ 
$^5$Dipartimento di Fisica, Universit\`a di Roma ``La Sapienza'', P.le Aldo Moro 5, 00143, Roma, Italy}

\begin{document}

\date{Accepted, Received; in original form }

\pagerange{\pageref{firstpage}--\pageref{lastpage}} \pubyear{2015}

\maketitle

\label{firstpage}

\begin{abstract}
The abundance of O in planetary nebulae (PNe) has been historically used as a
metallicity indicator of the interstellar medium (ISM) where they originated;
e.g., it has been widely used to study metallicity gradients in our Galaxy and
beyond. However, clear observational evidence for O self enrichment in
low-metallicity Galactic PNe with C-rich dust has been recently reported. Here
we report asymptotic giant branch (AGB) nucleosynthesis predictions for the
abundances of the CNO elements and helium in the metallicity range $Z_{\odot}/4
< Z < 2Z_{\odot}$. Our AGB models, with diffusive overshooting from all the
convective borders, predict that O is overproduced in low--Z low-mass 
($\sim$1$-$3 M$_{\odot}$) AGB stars and nicely reproduce the recent O
overabundances observed in C-rich dust PNe. This confirms that O is not always a
good proxy of the original ISM metallicity and another chemical elements such as
Cl or Ar should be used instead. The production of oxygen by low-mass stars
should be thus considered in galactic-evolution models.
\end{abstract}


\section{Introduction}
Stars in the mass range 1 M$_{\odot}$ $<$ M $<$ 8 M$_{\odot}$ evolve through the
asymptotic giant branch (AGB), just before they form planetary nebulae (PNe) and
end their lives as white dwarfs. AGB stars are supported by H burning, which is
periodically interrupted by shell He burning above the degenerate core, where
3$\alpha$ nucleosynthesis takes place; the so-called thermal pulses
\citep[e.g.][]{schw65}. The main processes of nucleosynthesis take place during
the thermally-pulsing (TP) AGB phase. The H and He burning shell products are
transported to the stars' surface via the third dredge-up (TDU) during the
TP-AGB, converting originally O-rich stars into C-rich ones. Also, the stellar
outer layers can be enriched in products of the so-called hot bottom burning
\citep[HBB, e.g.][]{SackmannBoothroyd1992,Mazzitelli1999} process for the more
massive (M$>$3-4 M$_{\odot}$) AGB stars \citep[e.g.][]{garcia07}, which avoids
the formation of C-rich stars. At the end of the AGB phase,  low-mass
($\sim$1.5$-$3-4 M$_{\odot}$) stars are predicted to be C-rich (C/O$>$1), while
more massive stars, because of HBB activation, would remain O-rich (C/O$<$1)
during the entire AGB evolution.

The theoretical modelling of AGB stars has been significantly improved in the
last years; the last generation of AGB models even including dust formation
\citep[e.g.][]{paperIV, nanni13}. There exist, however, important differences
among the results obtained from different AGB models. This is mainly due to our
poor knowledge of stellar convection and mass loss, which profoundly affect the
nucleosynthesis results obtained \citep{vd05a, vd05b, doherty14a, doherty14b}.
We are still far from a self--consistent and physically sound treatment of both
processes. The only way to make a significant progress is to compare theoretical
expectations with the  observational evidence.

The chemical composition of PNe proves an extremely useful and valuable tool to
constrain AGB models \citep{marigo03, marigo11, letizia09, ventura15}; the chemistry of PNe
is the outcome of the combination of the different processes that contribute to
alter the surface chemical composition during the whole AGB life. In addition,
PNe - because of their emission-line nature - are easily observed at very large
distances, and the gas chemical composition can be derived. Some chemical
elements (e.g. Ar and Cl) may remain practically unchanged in PNe, reflecting
the primordial composition of the interstellar medium (ISM) where their central
stars were born. Other elements like C and N (to a lesser extent O), however,
may be strongly modified during the previous AGB phase. The O abundance in PNe
has been widely used as a metallicity indicator of the ISM where they
originated; e.g., to derive metallicity gradients in our own Galaxy and other
nearby galaxies \citep[e.g.][]{Perinotto06,Maciel10,Stanghellini14,Richer15}.
This is because ``standard" (with no extra-mixing processes) nucleosynthesis
theoretical models of low- and intermediate-mass stars available in the
literature \citep[e.g.][]{karakas10} do not predict significant O enrichment (or
destruction) at near-solar metallicities during the AGB phase. 

However, by using high-quality optical spectra in conjunction with
the best available ionization correction factors (ICFs), \citet{gloria} have
very recently reported the first observational evidence of O self-enrichment (by
$\sim$0.3 dex) in Galactic PNe with C-rich dust; the expected outcome of
low-mass ($\sim$1.5$-$3 M$_{\odot}$) stars. As we mentioned above, the AGB
nucleosynthesis theoretical models with no extra-mixing processes do not predict
such O enrichment in low-mass stars. Models that include diffusive convective
overshooting \citep[e.g.][]{Herwig97,Marigo01}\footnote{We note that the
\citet{Pignatari13} models also seem to predict O production at the level
observed in C-rich dust PNe but these models have not been accepted for
publication yet.} predict a significant production of oxygen, even at solar
metallicities. This mixing process \citep[first introduced by][]{Herwig97} in
combination with an efficient TDU, may produce an increase of the O abundance
for low-mass ($\sim$1.5$-$3 M$_{\odot}$) stars. Rotation, magnetic fields, and
thermohaline mixing are another mechanisms, even less understood, that may cause
extra-mixing \citep[e.g.][]{karakas14}. Here we report self-consistent AGB
nucleosynthesis predictions (based on the ATON code) for the CNO elements and He
in the metallicity range $Z_{\odot}/4 < Z < 2Z_{\odot}$. Our AGB models predict
that O is overproduced in low-mass AGB stars and nicely reproduce the recent O
overabundances observed in C-rich dust PNe by \citet{gloria}. 

\section{The AGB ATON models}

The AGB models presented here are computed with the stellar evolution ATON code
\citep{mazzitelli89}. The numerical structure of ATON is given by
\citet{ventura98}, while \citet{ventura09} reported the most recent updates in
the code. A detailed description (i.e., numerical and physical input) of these
AGB ATON models and discussion in terms of the evolution of the surface
chemistry during the AGB phase - the role of mass and metallicity, the evolution
of CNO elements, etc.- will be presented in \citet{ventura16}. Here we
concentrate on the abundances of He, CNO elements and Cl (taken as a metallicity
indicator) at the end of the AGB phase and their comparison with the
corresponding abundances observed in C-rich dust PNe (see Section 4).

Briefly, the AGB ATON models are calculated using the following  physical
ingredients: (i) convection is modelled according to the full spectrum of
turbulence (FST) model \citep{cm91}. In regions unstable to convection, mixing
of chemical and nuclear burning are coupled by a diffusion--like equation
\citep{cloutmann76}. Overshoot of convective eddies into radiatively stable
regions follows an exponential decay of velocities from the border of the
convective zones (which is fixed via the Schwarzschild criterion) with the
e--folding distance of the decay given by $\zeta H_p$=0.002
\citep{paperIV}\footnote{The use of $\zeta=0.002$ mimiks overshoot from the base
of the convective envelope and from the borders of the convective shell forming
at the ignition of each thermal pulse \citep[see][for more details]{paperIV}.};
(ii) the \citet{blocker95} and \citet{wachter02} mass-loss prescriptions are
used for O- and C-rich AGB stars, respectively; (iii) the molecular opacities at
low temperatures ($<$$10^4$K) are calculated with the AESOPUS tool
\citep{marigo09}; this is especially important for the description of the C-rich
phase \citep{vm10}.

Table 1 summarizes the initial chemical composition of the AGB ATON models. They
cover all progenitor masses (i.e. $1~M_{\odot} \leq M \leq 8~M_{\odot}$) of
stars evolving through the AGB phase from subsolar to supersolar metallicity
($Z=4\times 10^{-3}$, $8\times 10^{-3}$, 0.018 and 0.04). The low-Z models
presented here, including the discussion of the evolutionary sequences, are
extensively illustrated in \citet{ventura14b} ($Z=4\times 10^{-3}$),
\citet{ventura13} ($Z=8\times10^{-3}$; M$\geq$3 M$_{\odot}$) and \citet{paperIV}
($Z=8\times10^{-3}$; M$\leq$3 M$_{\odot}$). The solar/supersolar metallicity
models have been calculated appositely for the present analysis \citep[see][for
more details]{ventura16}. An $\alpha-$enhancement $[\alpha/Fe]=+0.2$ is used
for the two lower metallicities, while the chemical mixture for the $Z=0.018$
and 0.04 models is solar--scaled. The solar composition from \citet{gs98} is
assumed. We note that the He abundance and the abundance ratios studied here are
consistent with similar models calculated with more recent solar abundances
\citep[e.g.][]{Asplund09}.

\section{Brief overview of the AGB ATON model predictions}

\subsection{Changes in the AGB chemistry: TDU and HBB}

The surface abundances of AGB stars are altered by TDU and HBB \citep[see
e.g.][]{karakas14}. The TDU is the mixing of nuclearly processed matter with the
external regions and mainly increases the surface C abundance (also O but to a
lesser extent). The HBB, however, is activated at the base of the convective
envelope for T$_{bce}$$\gtrsim$30 MK and favours CN nucleosynthesis (N
production); for T$_{bce}$$\gtrsim$80 MK, full CNO cycling is activated, which
further produces N at the expenses of C and O. The efficiency of TDU and HBB
mainly depends on the initial stellar mass and metallicity although the model
results are highly sensitive to the treatment of the convective borders and to
the adopted model for convection. 

Both TDU and HBB mechanisms are more efficient at lower metallicity (e.g. the
HBB temperatures decrease with increasing metallicity and the high-Z
models experience less extended nucleosynthesis). We may distinguish three main
cases depending on the progenitor masses: (i) stars with progenitor masses
higher than M$_{up}$ experience strong HBB at T$_{bce}$$\gtrsim$80 MK and
limited TDU; He (mostly due to the second dredge-up, SDU) and N are enhanced
with C and O being destroyed. The threshold mass, $M_{up}$, increases with
decreasing metallicity (Table 1); (ii) stars with progenitor mass between
$M_{HBB}$ and $M_{up}$ also experience HBB but TDU is more efficient than in
their higher mass counterparts; the lower metallicity models experience a few
late TDU episodes that may convert the star into C-rich. In all models N is
enhanced together with some He enrichment (again from the SDU) but C and O may
be created or destroyed depending on the dominant mechanism (TDU or HBB). The
mass limit for HBB activation, $M_{HBB}$, increases with increasing metallicity
(Table 1); (iii) stars with progenitor mass lower than $M_{HBB}$ do not
experience HBB because T$_{bce}$$<$30 MK and TDU dominates their chemistry,
mainly increasing the surface C content (and O at a smaller extent). The minimum
mass to form C-rich stars\footnote{Stars with initial mass lower than $M_C$ are
not converted to C-rich (C/O$>$1) because their chemistry is only affected by
the first dredge-up during the red giant branch.}, $M_C$, decreases with
decreasing metallicity (Table 1) because at lower metallicity TDU is more
efficient and less carbon is needed to attain the C-star stage.

\begin{table}
\begin{center}
\caption{Chemical properties of AGB ATON models} 
\begin{tabular}{c|c|c|c|c|c|}
\hline
Z & Y & $[\alpha /Fe]$ & $M_C$ & $M_{HBB}$ & $M_{up}$  \\ 
\hline
$4\times 10^{-3}$  & 0.25   &  $+$0.2   &  1.1        &  3.5  &  6.0  \\ 
$8\times 10^{-3}$  & 0.26   &  $+$0.2   &  1.2        &  3.5  &  6.0  \\ 
0.018		   & 0.28   &     0.0   &  1.4        &  3.5  &  5.5  \\ 
0.04		   & 0.30   &     0.0   &  $\dots$    &  4.0  &  4.0  \\ 
\hline
\end{tabular}
\end{center}
\label{tabmod}
\end{table}

\begin{figure*}
\begin{minipage}{0.4\textwidth}
\resizebox{1.\hsize}{!}{\includegraphics{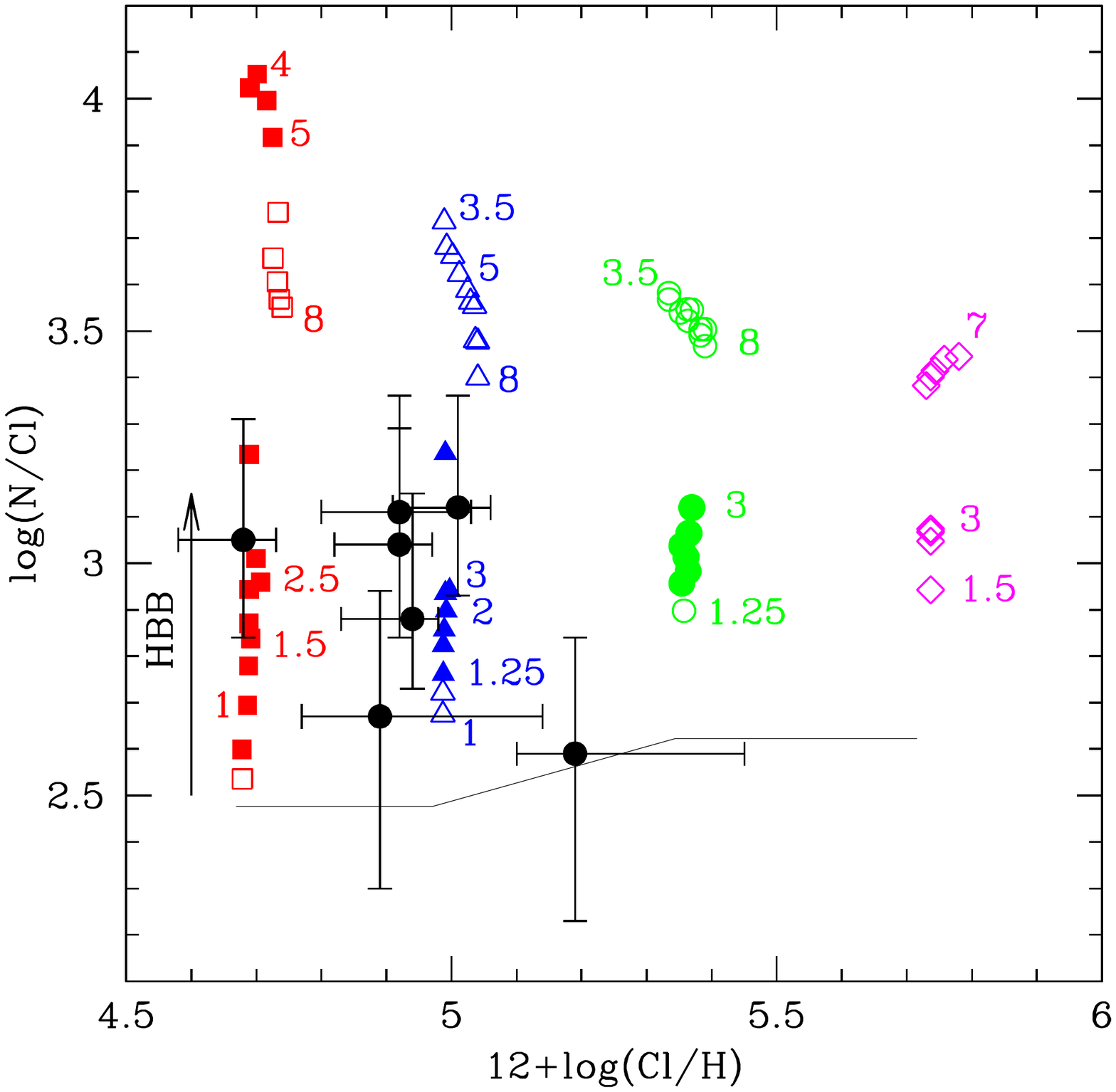}}
\end{minipage}
\begin{minipage}{0.4\textwidth}
\resizebox{1.\hsize}{!}{\includegraphics{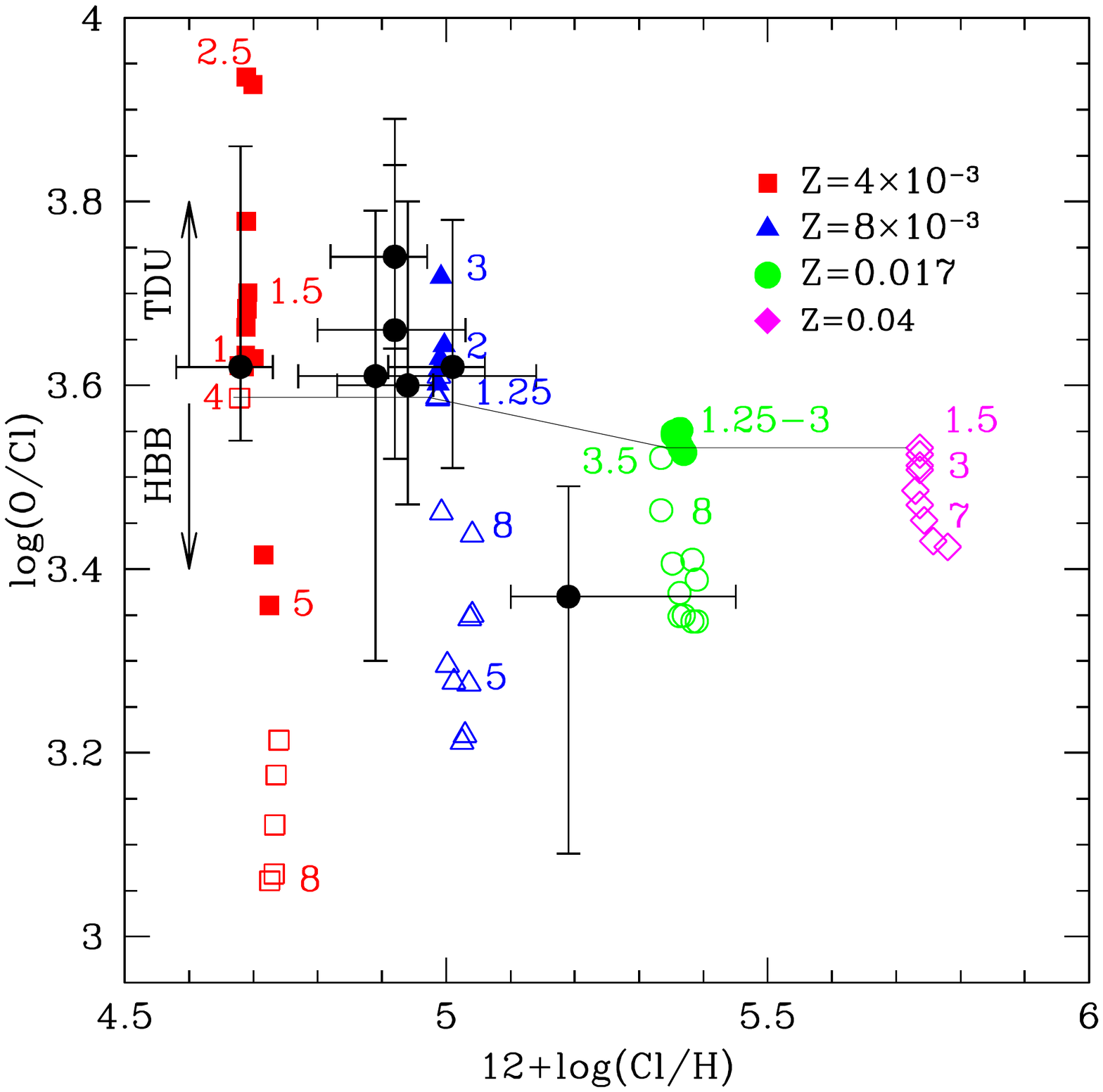}}
\end{minipage}
\vskip-70pt
\begin{minipage}{0.4\textwidth}
\resizebox{1.\hsize}{!}{\includegraphics{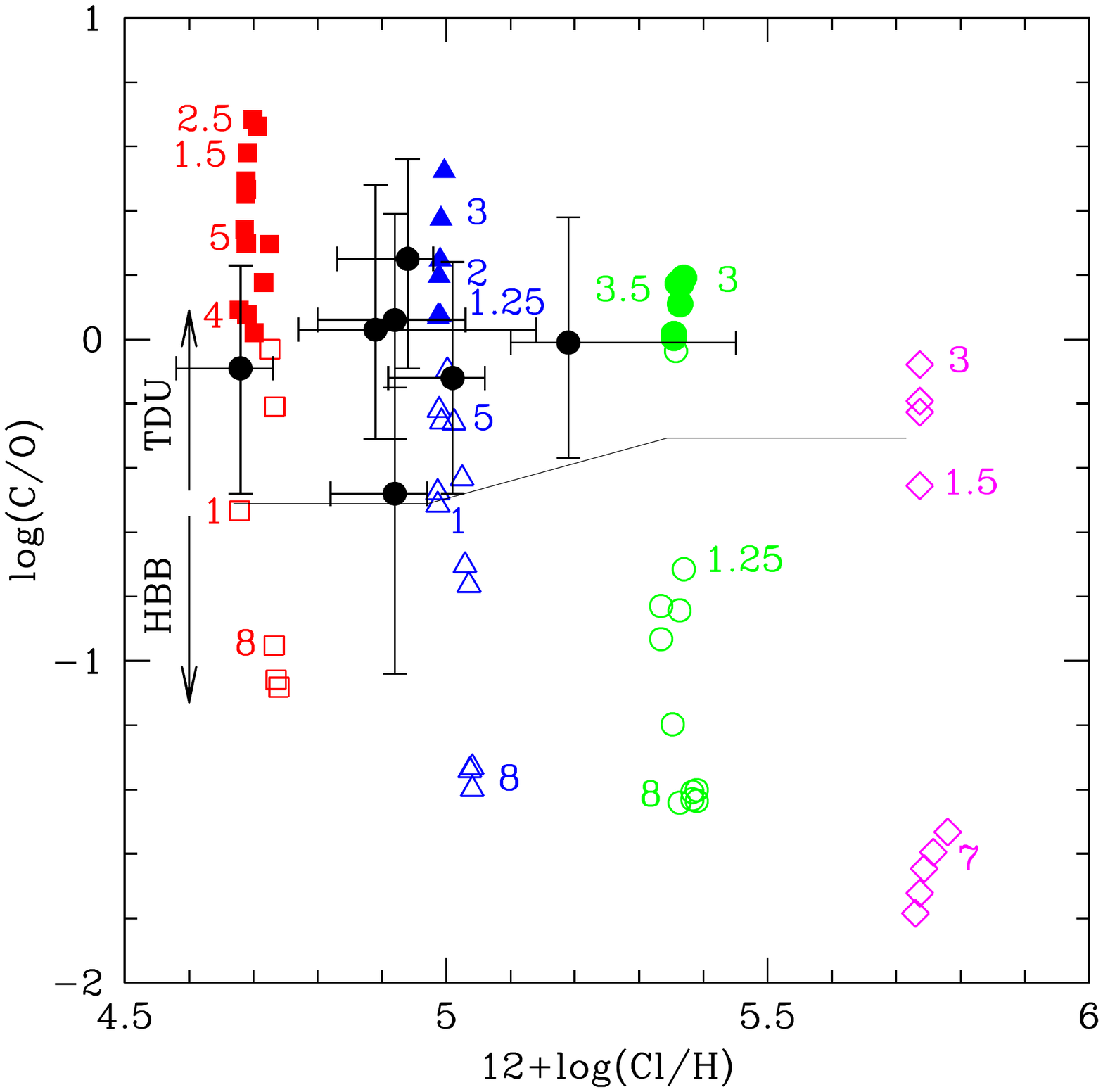}}
\end{minipage}
\begin{minipage}{0.4\textwidth}
\resizebox{1.\hsize}{!}{\includegraphics{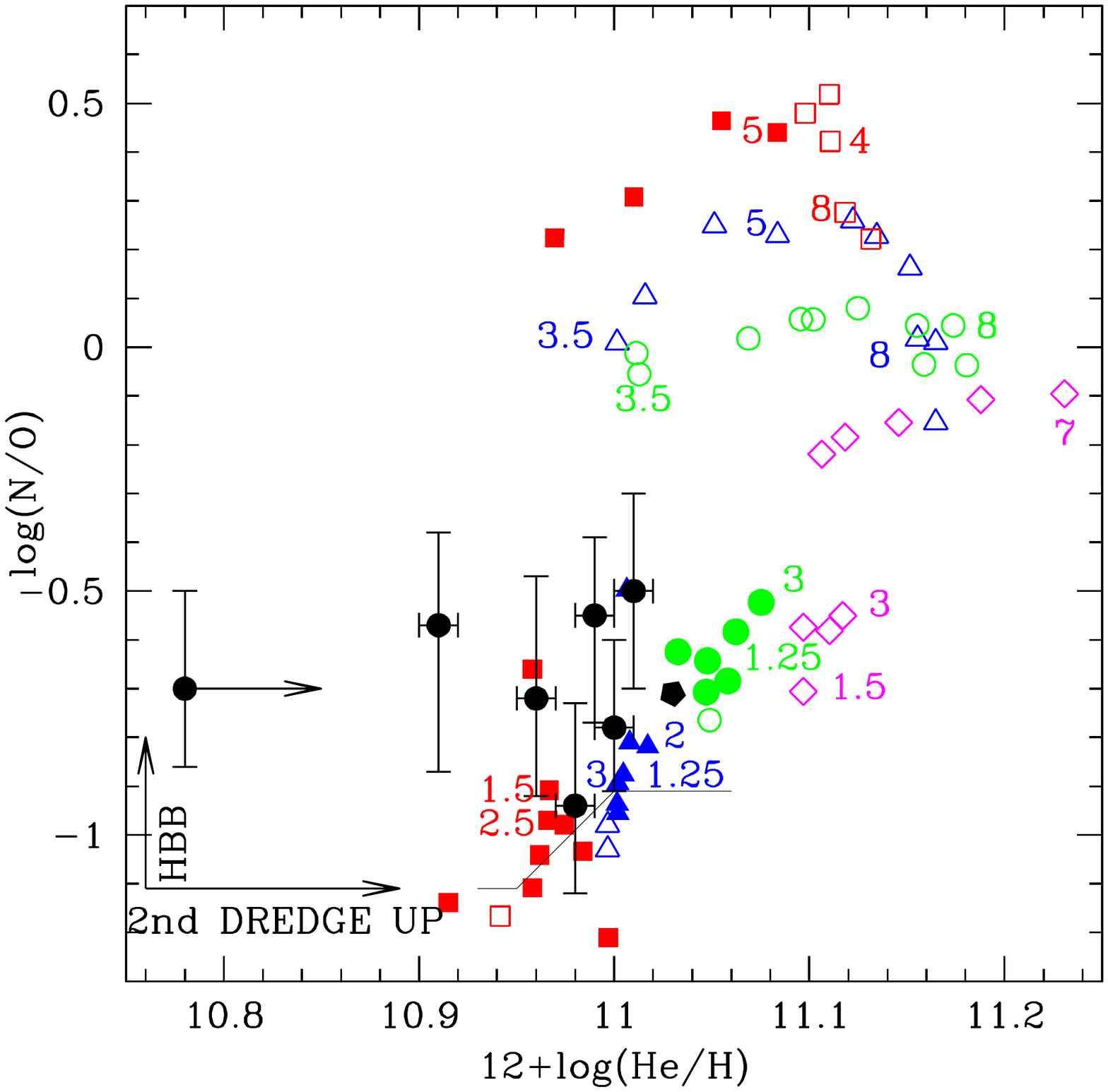}}
\end{minipage}
\vskip-40pt
\caption{Chemical abundances in Galactic C-rich dust PNe (black dots,
Delgado-Inglada et al. 2015) vs the ATON AGB model predictions for different
masses (a few relevant masses are marked) and metallicities: Cl vs. N/Cl
(top, left), Cl vs. O/Cl (top, right), Cl vs. C/O  (bottom, left), He vs. N/O
(bottom, right). The thin, solid lines indicate the assumed initial abundances,
while the arrows indicate the qualitative effect of HHB, SDU and TDU.
Filled and open symbols correspond to C- and O-rich stars, respectively.
The black pentagon (bottom-right panel) is the median abundance of Galactic
C-rich dust PNe measured by \citet{garcia14}.}
\label{fpne}
\end{figure*}

\subsection{Surface chemistry at the end of the AGB}

The temporal evolution of the surface chemistry of the AGB ATON models allows to
calculate the yields of the different chemical elements \citep[see][for a full
discussion on this]{ventura16}. However, the interpretation of nebular
abundances in PNe requires the surface mass fractions (or surface abundances) of
the various chemical elements at the end of the AGB phase.

The final HeCNOCl surface abundances of the AGB ATON models used in the present
analysis are shown in the four panels of Fig.\ref{fpne}. We show also the 
chlorine abundance together with He and the CNO elements because Cl is taken as
a good metallicity indicator by \citet{gloria}. Chlorine is not expected to
undergo any processing during the AGB evolution, remaining constant during the
whole life of the star. The surface Cl abundance is therefore representative of
the gas from which the star was formed. It is to be noted here that for
consistency with the AGB ATON models, we consider the solar Cl abundance of $12
+ log(Cl/H)$=5.50 \citep{gs98}\footnote{This also corresponds to the most recent
solar Cl abundance as determined by \citet{Asplund09}.}, while \citet{gloria}
considered a lower solar Cl abundance of $12 + log(Cl/H)$=5.26, as reccomended
by \citet{Lodders03}. This means that all C-rich dust PNe in their sample are of
subsolar metallicity here\footnote{This is in perfect agreement with
\citet{garcia14} who derived homogeneously the Cl and Ar abundances in a large
sample of C-rich dust Galactic PNe.}, while \citet{gloria} considered the few
most Cl-rich objects as near-solar metallicity PNe.

Fig.\ref{fpne} shows the final N and O abundances and the C/O and N/O ratios
versus Cl and/or He in our AGB ATON models compared with the observations of
C-rich dust PNe by \citet{gloria}\footnote{The abundances of C, Cl, He and O
were obtained using the ICFs derived by \citet{gloria14} for PNe. The N
abundances were computed using the classical expression N/O=N$^{+}$/O$^{+}$ that
seems to work better \citep[see the discussion in][]{gloria14}.}. Lower
metallicity models cover a wider range of N and O compared to their solar and
supersolar counterparts (see the top panels in Fig.1). The reason for this is
twofold: a) in the massive AGB domain, lower-Z models experience a stronger HBB,
which  favours lower O and higher N; b) in the low-mass regime the efficiency of
TDU is higher the lower is the metallicity, which favours the O increase and the
production of primary N (see also Section 3.1). In the ATON models, the
individual C and O abundances are extremely sensitive to the metallicity, but
the C/O ratio is similar for the four metallicities; the highest C/O ratio (over
a factor 10 higher than in the initial mixture) is found in the $Z=4\times
10^{-3}$ models, as expected. The interpretation of the N/O vs. He plane
(right--bottom  panel of Fig.\ref{fpne}) is less straightforward. The N/O ratio
is generally used as a HBB strength indicator, while He is connected to the
inwards penetration of the convective envelope during the SDU. The SDU
efficiency (which increases He at the stellar surface) mainly increases with
increasing stellar mass \citep{ventura10}. Thus, the model predictions are
distributed dychotomically in the N/O vs. He plane. Low-mass AGB stars define an
approximately vertical sequence at constant He (no SDU is expected below $\sim
4~M_{\odot}$ and variable N/O ($\lesssim$0.3)), while higher mass HBB stars are
spread in the high N/O and He region. 

\section{Understanding O-enrichment in C-rich dust PNe}

The C-rich dust PNe sample (7 objects) of \citet{gloria} span a range of
metallicities, as traced by the Cl content, extending over one order of
magnitude. Most C-rich dust PNe (5 sources), however, display subsolar Cl
abundances of $12 + log(Cl/H)$$\sim$4.9$-$5.0 dex. The other two objects are Hu
2$-$1 ($12 + log(Cl/H)$=4.68 dex) and NGC 6826 ($12 + log(Cl/H)$=5.19 dex). The
more metal-rich PN (NGC 6826) is suspected to have a binary companion
\citep{Mendez89}, which might produce its carbon enrichment, and we do not
consider it in the subsequent discussion. 

Fig. \ref{fpne} shows that the chemical composition of 5 out of the 6
low-metallicity C--rich dust PNe is nicely reproduced by our low--mass 
($\sim$1$-$3 M$_{\odot}$) models of metallicity $Z=8\times 10^{-3}$. The
theoretical predictions account for the slight O enhancement (by $\sim$0.2$-$0.3
dex) observed in these PNe, but also for the N and He abundances as well as for
the C/O and N/O ratios. The chemical composition of the lowest metallicity PN
(Hu 2$-$1) is also nicely reproduced by our lowest metallicity models
($Z=4\times 10^{-3}$), with the exception of He that is more abundant than
predicted. Higher mass (M$>$3 M$_{\odot}$) models experience HBB (marked
with open symbols in Fig. 1) with the consequent decrease of C, leading to
C/O$<$1. These models do not predict the observed O-enrichment neither the N/Cl
and N/O ratios (and He abundances). Therefore we do not expect C-rich dust PNe
arise from high-mass progenitors. As far as we know, our models are the first
ones giving a self-consistent explanation for the O-enrichment observed in
C-rich dust PNe. Also, our AGB models, when coupled with dust formation, predict
these objects to be rich in C dust \citep[see e.g.][]{ventura15}, as observed.  

In short, the main difference between the ATON models and another published AGB
models\footnote{The \citet{Marigo01} AGB models are based on synthetic
computations, which renders difficult a straight comparison with our ATON
results, rather based on a self-consistent description of the whole stellar
structure.} is that ATON assumes extra-mixing (diffusive overshooting) from all
the convective borders, including the bottom of the convective shell, while
other models use only overshoot from the base of the convective envelope. When
the extra-mixing from the convective shell is used, the pulse is stronger (which
also leads to a more penetrating TDU) and convection reaches more internal
layers, where the oxygen content is higher. \citet{Herwig97} also assume a tiny
overshoot (comparable to ATON) at the base of the pulse driven convective zone
but their overshoot from the base of the convective zone is much higher than in
ATON. Unfortunately, \citet{Herwig97} did not report the final O surface
abundances predicted by their 3 M$_{\odot}$ solar metallicity AGB models. Their
treatment of boundaries of convective regions leads to intershell abundances of
typically ($^4$He/$^{12}$C/$^{16}$O)=(23/50/25) (compared to (70/26/1) in the
standard treatment), while intershell abundances of (50/37/6) are obtained in
our ATON model (2M$_{\odot}$, Z=0.004) undergoing the largest O enrichment.
Thus, we would expect the \citet{Herwig97} models to predict more surface O than
observed in C-rich dust PNe. 

The characterisation of the formation epoch and of the progenitor masses is
hampered by uncertainties associated to the individual abundances, particularly
of the C/O ratio. In addition, the formation epoch estimates are
model-dependent; e.g., the ATON evolutionary timescales are generally much
shorter than the Karakas (2010) models. We may conclude, conservatively 
and according to ATON, that these C-rich dust PNe descend from stars of initial
mass in the range $1.5~M_{\odot} < M < 3~M_{\odot}$, formed between $\sim$400
Myr and 2 Gyr ago. If we assume the recommended C/O values given by
\citet{gloria}, we may rule out progenitors more massive than $2~M_{\odot}$;
indeed models of mass $M \geq 2.5~M_{\odot}$ undergo the most noticeable
variation in the surface chemical composition, reaching C/O$\sim$3, which is
significantly higher than the maximum C/O observed in these C-rich dust PNe
(i.e. C/O$\sim$1.8 in PN IC 418). This upper limit to the observed C/O in C-rich
dust PNe, if confirmed, would suggest an even narrower range of progenitor
masses, namely $1.5~M_{\odot} \leq M \leq 2~M_{\odot}$, with formation epochs
extending from $\sim$1 to 2 Gyr ago. This finding is in agreement with recent
results on similar low-metallicity C-rich dust PNe in the Large Magellanic Cloud
\citep{ventura15}, which show C/O ratios below 2 and identical C-rich dust
features in their mid-IR {\it Spitzer} spectra \citep{letizia07}.

In summary, our AGB models with extra-mixing (diffusive overshooting) from
all the convective borders (including the bottom of the convective shell)
predict that O is overproduced in low--Z low-mass ($\sim$1$-$3
M$_{\odot}$) AGB stars and nicely reproduce the O overabundances observed in
C-rich dust PNe as well as their N and He abundances and CNO abundance ratios.
This shows that O is not always a good proxy of the original ISM metallicity and
another chemical elements such as Cl or Ar should be used instead. Oxygen
production by low-mass stars should be thus considered in Galactic chemical
evolution models.

\section*{Acknowledgments}

D.A.G.H. was funded by the Ram\'on y Cajal fellowship number RYC$-$2013$-$14182
and he acknowledges support provided by the Spanish Ministry of Economy and
Competitiveness (MINECO) under grant AYA$-$2014$-$58082-P. P.V. was supported by
PRIN MIUR 2011 `The Chemical and Dynamical Evolution of the Milky Way and Local
Group Galaxies' (PI: F. Matteucci), prot. 2010LY5N2T. G.D.I. acknowledges
support from the Mexican CONACYT grant CB$-$2014$-$241732.

\end{document}